\documentclass[twocolumn,prl,preprintnumbers,superscriptaddress,amsmath,amssymb,english]{revtex4}
\usepackage{amsmath,amssymb,amsfonts,mathrsfs}
\usepackage{amsthm}
\usepackage{CJK}
\usepackage{bm}
\usepackage{lipsum}
\usepackage{babel}
\usepackage{graphicx}
\allowdisplaybreaks
\usepackage{makecell}
\usepackage{booktabs}
\usepackage{multirow}
\usepackage{gensymb}
\usepackage{rotating}
\usepackage{float}
\usepackage{braket}
\usepackage[breaklinks]{hyperref}
\usepackage{color}
\hypersetup{colorlinks=true, linkcolor=blue, citecolor=blue, filecolor=blue, urlcolor=blue}
\UseRawInputEncoding

\begin{document}

\title{Floquet Engineering of Nonequilibrium Valley-Polarized Quantum Anomalous Hall Effect with Tunable Chern Number}

\author{Fangyang Zhan}
\affiliation{Institute for Structure and Function $\&$ Department of Physics $\&$ Chongqing Key Laboratory for Strongly Coupled Physics, Chongqing University, Chongqing 400044, P. R. China}

\author{Junjie Zeng}
\affiliation{Institute for Structure and Function $\&$ Department of Physics $\&$ Chongqing Key Laboratory for Strongly Coupled Physics, Chongqing University, Chongqing 400044, P. R. China}

\author{Zhuo Chen}
\affiliation{Institute for Structure and Function $\&$ Department of Physics $\&$ Chongqing Key Laboratory for Strongly Coupled Physics, Chongqing University, Chongqing 400044, P. R. China}

\author{Xin Jin}
\affiliation{Institute for Structure and Function $\&$ Department of Physics $\&$ Chongqing Key Laboratory for Strongly Coupled Physics, Chongqing University, Chongqing 400044, P. R. China}

\author{Jing Fan}
\affiliation{Center for Computational Science and Engineering, Southern University of Science and Technology, Shenzhen 518055, P. R. China}

\author{Tingyong Chen}
\email[]{chenty@sustech.edu.cn}
\affiliation{Shenzhen Insitute for Quantum Science and Engineering, Southern University of Science and Technology, Shenzhen 518055, P. R. China}

\author{Rui Wang}
\email[]{rcwang@cqu.edu.cn}
\affiliation{Institute for Structure and Function $\&$ Department of Physics $\&$ Chongqing Key Laboratory for Strongly Coupled Physics, Chongqing University, Chongqing 400044, P. R. China}
\affiliation{Center for Computational Science and Engineering, Southern University of Science and Technology, Shenzhen 518055, P. R. China}
\affiliation{Center of Quantum materials and devices, Chongqing University, Chongqing 400044, P. R. China}

\begin{abstract}
Numerous attempts have been made so far to explore the quantum anomalous Hall effect (QAHE), but the ultralow observed temperature strongly hinders its practical applications. Hence, it is of great interest to go beyond the existing paradigm of QAHE. Here, we propose that Floquet engineering offers a strategy to realize the QAHE via hybridization of Floquet-Bloch bands. Based on first-principles calculations and Floquet theorem, we unveil that nonequilibrium valley-polarized QAHE (VP-QAHE), independent of magnetic orders, is widely present in ferromagnetic and nonmagnetic members of two-dimensional family materials $M$Si$_2$$Z_4$ ($M$ = Mo, W, V; $Z$ = N, P, As) by irradiating circularly polarized light (CPL). Remarkably, by tuning the frequency, intensity, and handedness of incident CPL, the Chern number of VP-QAHE is highly tunable and up to $\mathcal{C}=\pm 4$. We reveal that such Chern number tunable VP-QAHE attributes to light-induced trigonal warping and multiple band inversion at different valleys. The valley-resolved chiral edge states and quantized plateau of Hall conductance, which facilitates the experimental measurement, are visible inside the global band gap. Our work not only establishes Floquet Engineering of nonequilibrium VP-QAHE with tunable Chern number in realistic materials, but also provides a promising avenue to explore emergent topological phases under light irradiation.
\end{abstract}


%Use showkeys class option if keyword

\maketitle
The quantum anomalous Hall effect (QAHE), which was first theoretically proposed by Haldane in 1988 \cite{PhysRevLett.61.2015}, is an attractive topological quantum phenomenon that exhibits quantized Hall conductance $\sigma_{xy}=\mathcal{C}\frac{e^2}{h}$, where $\mathcal{C}$ is the first Chern number \cite{PhysRevLett.49.405}. The QAHE can be considered as a zero magnetic field counterpart of the quantum Hall effect and usually occurs in two-dimensional (2D) ferromagnetic (FM) insulators featured by a nontrivial bulk gap, termed as quantum anomalous Hall (QAH) insulators or Chern insulators. The spin-polarized dissipationless chiral edge channels of QAHE offer a fascinating avenue for designing future electronic and spintronic devices with low-power consumption. Meanwhile, the interplay between band topology and FM order in QAH insulators would provide a fertile platform to explore exotic quantum phenomena, for instance, topological axion electrodynamics and unconventional superconductivity \cite{PhysRevB.82.184516,PhysRevLett.120.056801}. Therefore, seeking the QAHE in realistic systems %, which can be further potential to practical application, %, which can easily be fabricated in the laboratory%and further potential to practical application
is a significant topic of intense recent studies. In the past decade, though there have been numerous theoretical proposals to predict the QAHE \cite{PhysRevLett.101.146802,science.1187485,PhysRevLett.108.056802,PhysRevLett.110.196801,PhysRevLett.113.256401,PhysRevB.95.125430,PhysRevLett.112.116404,PhysRevB.98.245127,sciadv.aaw5685,PhysRevLett.122.107202,PhysRevLett.128.026402,PhysRevLett.129.036801}, its experimental realization is quite limited. Up to now, the QAHE has only been observed in magnetically doped topological insulators (TIs) \cite{Chang167,PhysRevLett.113.137201,2014Trajectory,Cui2015High,2020Tuning}, intrinsically magnetic TI MnBi$_2$Te$_4$ films \cite{science.aax8156}, twisted bilayer graphene \cite{science.aay5533}, and MoTe$_2$/WSe$_2$ heterobilayers \cite{2021Quantum}.

As is well-known, the critical issue in the field of QAHE is improving the observation temperature, which requires the coexistence of robust FM order (i.e., strong FM exchange) and a sizable nontrivial band gap that is often opened by the spin-orbital coupling (SOC) effect. But in fact, the 2D ferromagnetism tends to arise in metallic systems composed of light elements, while the strong SOC effect is inclined to associate with heavy elements. As a result, most 2D FM materials are metallic, and although several 2D FM semiconductors have recently been fabricated in the laboratory \cite{Huang2017,Gong2017}, few of them possesses a nontrivial band gap.
%The QAH insulators reported so far have weak FM coupling, which poses great challenges for the field of QAHE related electronics.
To solve these problems, one would go beyond the existing paradigm of QAHE.

Recently, Floquet engineering (i.e., periodic driving) provides a wide range of pathways to dynamically manipulate topological states, leading to emergent nonequilibrium phases that are otherwise not possible in equilibrium cases \cite{lindner2011floquet,mciver2020light, science.1239834,2016Selective,PhysRevLett.120.237403,Light2022,PhysRevB.84.235108, PhysRevB.90.115423, PhysRevB.79.081406, PhysRevLett.110.026603}.
More strikingly, irradiation of circularly polarized light (CPL) can break the time-reversal symmetry (TRS) \cite{PhysRevB.84.235108,PhysRevB.90.115423,PhysRevB.79.081406, PhysRevLett.110.026603}. It was experimentally demonstrated that,  under irradiation of CPL,  a light-induced anomalous Hall effect (AHE) was observed in graphene \cite{mciver2020light} and the surface Dirac cone of three-dimensional topological insulator was gapped \cite{science.1239834}. Consequently, the further exploration of light-induced QAHE  has brewed an attractive topic \cite{PhysRevB.103.195146, PhysRevB.105.035103, PhysRevB.105.L081115}. Nonetheless, the attempts made so far have mainly focused on theoretical models or corollaries, and material candidates that can realize light-induced QAHE are largely unexplored. %Hence, to facilitate the experimental observation,
In particular, it is unclear how to design QAHE via utilizing Floquet engineering in realistic materials.

\begin{figure}
    \centering
    \includegraphics[scale=0.97]{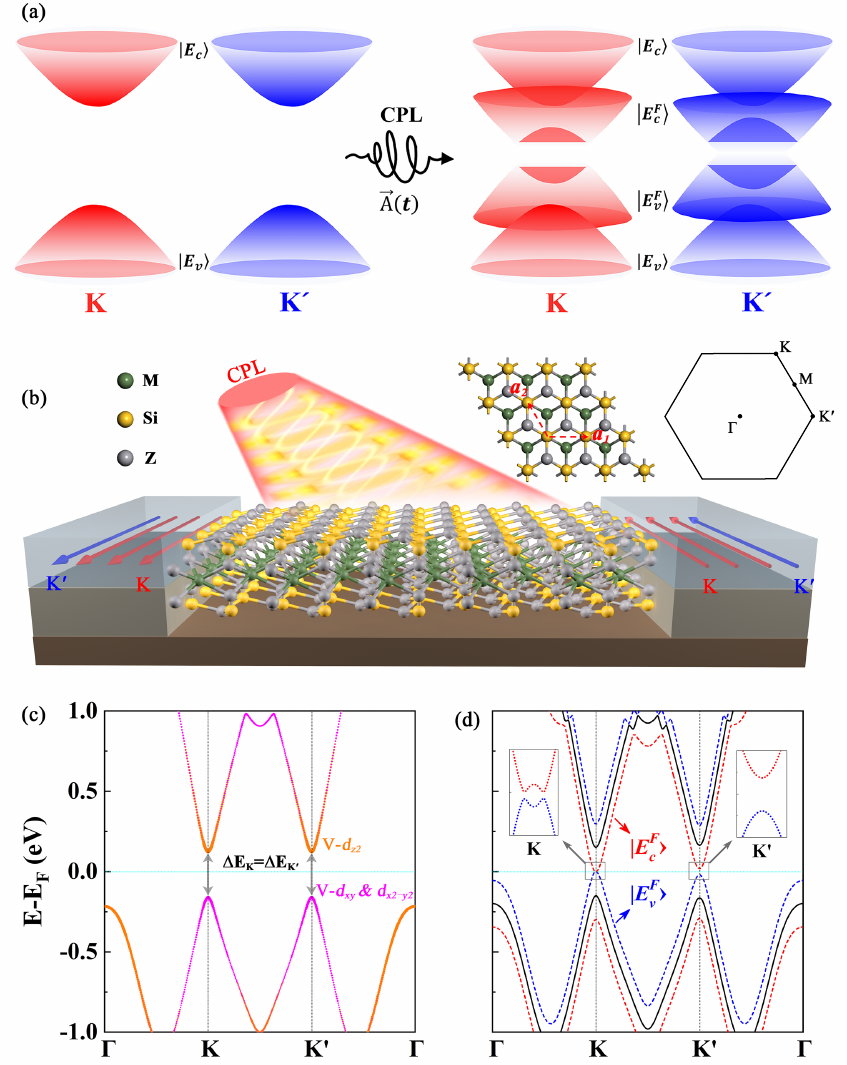}
    \caption{(a) Band diagram illustrating valley-dependent band inversion via hybridization of Floquet-Bloch bands (i.e., $E_c^{F}$ and $E_v^{F}$) originated from irradiation of CPL $\mathbf{A}(t)$. (b) Schematic diagram of light-induced chiral edge channels of high Chern number VP-QAHE. The insets give the top view of lattice structures of 2D $M$Si$_2$Z$_4$ family materials and hexagonal Brillouin zone (BZ). (c) The orbital-resolved electronic band structure of monolayer VSi$_2$N$_4$ including SOC with the magnetization along the $x$-axis. The components of V $d_{xy}+d_{x^{2}-y^{2}}$ and V $d_{z^2}$ orbitals are proportional to the width of the magenta and orange lines, respectively. (d) The photon-dressed band structures of VSi$_2$N$_4$ subject to left-handed CPL with a certain light intensity and frequency (i.e., $\hbar \omega=0.145$ eV and $eA/\hbar=0.028$ {\AA}$^{-1}$). The black solid lines represent the equilibrium bands. The blue and red dashed lines represent Floquet-Bloch bands created by absorbtion and emission of photons, respectively. The insets indicate that two Floquet-Bloch bands invert at the $K$ point and preserve the trivial band gap at the $K'$ point.
    \label{FIG1}}
\end{figure}

In this Letter, we point out that Floquet engineering with CPL offers a fascinating strategy to realize the QAHE with great tunability. Due to spontaneous TRS breaking by irradiating CPL, we emphasize that our proposal can generally be applied in magnetic and nonmagnetic (NM) systems. The key idea of our strategy is utilizing the momentum- and/or spin-resolved features of photon-dressed Floquet-Bloch states to cause band inversion around specific local extrema (i.e., valleys) of valence and conduction bands. As a consequence, the QAHE related to valley-polarization (i.e., VP-QAHE) can be achieved by tuning the intensity or frequency of incident CPL. By the way, the VP-QAHE belongs to a special type of QAHE, which has stimulated extensive interest due to its combination of valleytronics and spintronics with band topology \cite{PhysRevLett.108.196802,PhysRevLett.112.106802,PhysRevLett.119.046403,PhysRevLett.127.116402,PhysRevB.105.L081115}. As a typical representative,
%The most representative materials of studying valleytronics crystallize in a 2D hexagonal lattice \cite{}. Hence,
we consider the evolution of band structures around two inequivalent valleys $K$ and $K'$ of a hexagonal Brillouin zone (BZ) to elucidate the formation process of VP-QAHE under irradiation of CPL. As illustrated in Fig. \ref{FIG1}(a), we start from a trivial insulator with direct band gaps at two valleys $K$ and $K'$ in the absence of light irradiation. Under irradiation of CPL, periodic driving gives rise to Floquet-Bloch bands, and then two certain bands [i.e., labeled as $E_c^{F}$ and $E_v^{F}$ in Fig. \ref{FIG1}(a)] move close to the Fermi level via the optical Stark effect \cite{2015Valley,PhysRevB.97.045307,nanolett.6b04419,science.aal2241}. With the lifting of valley degeneracies under light irradiation \cite{2015Valley}, the valley-resolved topological phase transition accompanied by band gap closing and reopening may occur [see the left panel of Fig. \ref{FIG1}(a)]. Notably, the nontrivial band gap here is derived from light-induced band hybridization instead of SOC.

Next, we demonstrate that light-induced VP-QAHE can be realized in 2D $M$Si$_2$$Z_4$ ($M$ = Mo, W, V; $Z$ = N, P, As) family materials subject to CPL. The $M$Si$_2$$Z_4$ family are a new class of 2D layered materials \cite{2021Intercalated,science.abb7023,PhysRevB.102.235435,PhysRevB.104.L201112,PhysRevB.104.075421}. This family of materials, including magnetic and NM members, host excellent stability \cite{science.abb7023}, and especially MoSi$_2$N$_4$ and WSi$_2$N$_4$ were successfully synthesized in experiments \cite{science.abb7023}. Based on first-principles calculations and Floquet theorem, we show that light-induced VP-QAHE can widely arise in the FM and NM semiconductor members in the $M$Si$_2$$Z_4$  family. Remarkably, we find that irradiation of CPL can %lifts the degeneracies of two valleys and thereby
give rise to exotic trigonal warping at one valley, resulting in the VP-QAHE with highly tunable Chern number, which is up to $\mathcal{C}=\pm 4$ [see Fig. \ref{FIG1}(b)]. The high Chern number QAHE is more favorable for dissipationless device applications due to its more chiral edge channels \cite{PhysRevLett.115.057206,PhysRevLett.111.136801}. It is worth noting that such high Chern number QAHE has only been investigated in static systems and not been reported in dynamic systems with periodic driving. In the main text, we take the FM VSi$_2$N$_4$ monolayer as an example to depict Floquet engineering of Chern number tunable VP-QAHE. The results of other candidate materials are included in the Supplemental Material (SM) \cite{SM}.

To investigate the topological phase transition in 2D $M$Si$_2$Z$_4$ family materials under periodic driving, we carried out first-principles calculations using the Vienna \textit{ab initio} simulation package \cite{PhysRev.140.A1133,PhysRevB.54.11169} to obtain electronic band structures with the basis of plane waves (see details in the SM \cite{SM}). Then the Wannier-function-based tight-binding (WFTB) Hamiltonian was constructed as implemented in Wannier90 package \cite{Mostofi2014}. By applying time-periodic and space-homogeneous CPL [see Fig. \ref{FIG1}(b)], the photon-dressed electronic band structures were calculated from the WFTB Hamiltonian combining with Floquet theorem \cite{PhysRevB.102.201105}. Here, we do not diagonalize the infinite dimensional Floquet Hamiltonian, but rather restrict the Floquet bands to first-order sidebands. The computed approach in details was shown in the SM \cite{SM}.

Without light irradiation, monolayer VSi$_2$N$_4$ is a FM semiconductor \cite{2021Intercalated,PhysRevB.103.085421,Sign-reversible}. To determine its easy magnetization axis, we have carried out total-energy calculations with different magnetic configurations in the presence of SOC. The results indicate that VSi$_2$N$_4$ favors to the in-plane magnetization and the magnetic anisotropy energy in the 2D plane is negligible, consistent with the previous studies \cite{PhysRevB.103.085421}. As shown in Fig. \ref{FIG1}(c), we plot orbital-resolved electronic band structures along high-symmetry directions of monolayer VSi$_2$N$_4$ including SOC with the magnetization along the $x$-axis. The bands, with the valence band and conduction band respectively contributed by $d_{xy}$ $\&$ $d_{x^{2}-y^{2}}$ and $d_{z^2}$ orbitals of V atoms, depict a trivial semiconducting feature with the valley degeneracy, i.e., band gaps $\Delta_{K}=\Delta_{K'}$. Besides, since the magnetization direction is possibly dependent on environmental conditions, we also give some brief discussion on the cases with magnetization along out-of-plane directions (see the Fig. S1 in the SM \cite{SM}). We can see that the change of magnetization only slightly lifts valley degeneracy but does not affect band topology.

\begin{figure}
    \centering
    \includegraphics[scale=0.99]{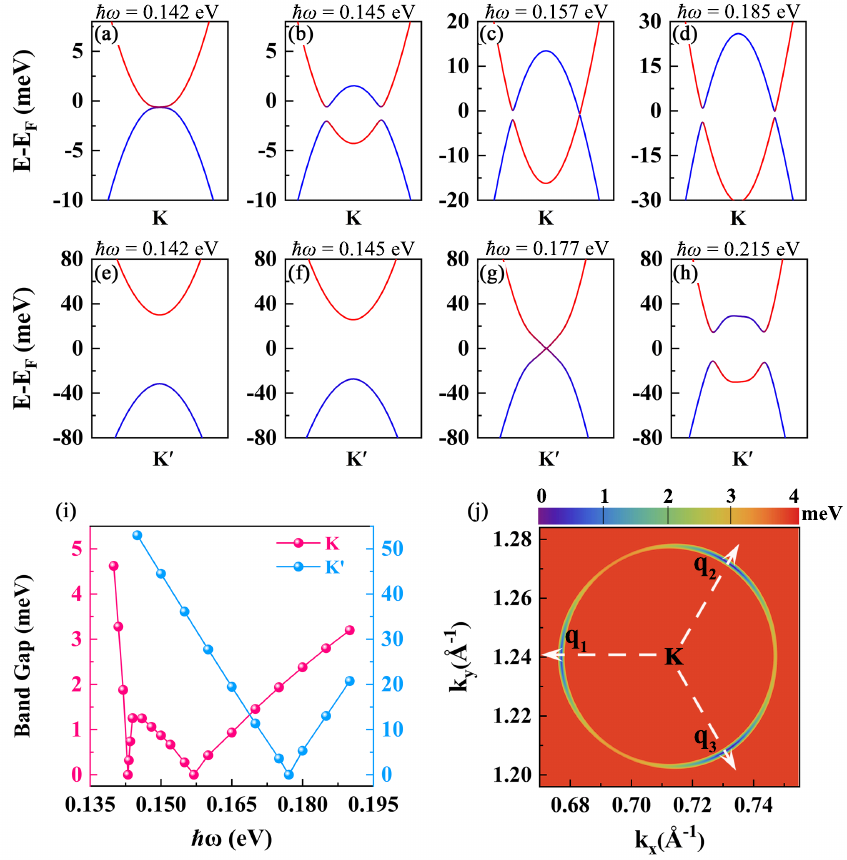}
    \caption{The evolution of Floquet-Bloch bands of VSi$_2$N$_4$ around the [(a)-(d)] $K$ and [(e)-(h)] $K'$ valleys with different photon energies $\hbar \omega$ to reveal the topological phase transitions. (i) The variation of band gaps around the $K$ and $K'$ valleys as a function of $\hbar \omega$. (j) The energy difference map between Floquet-Bloch bands $E_c^{F}$ and $E_v^{F}$ in the vicinity of $K$ valley with a critical photon energy $\hbar \omega=0.157$ eV. The light intensity is fixed at $eA/\hbar$ = 0.032 {\AA}$^{-1}$.
     \label{FIG2}}
\end{figure}

Under light irradiation, we choose CPL with a time-dependent vector potential of $\mathbf{A}(t)= A[\cos (\omega t ), \pm \eta \sin (\omega t ), 0]$, where $\eta=+1 (-1)$ denotes left- (right-) handed CPL, $A$ is the amplitude, and $\omega$ is the frequency. Thus, $\hbar \omega$ and $eA/\hbar$ represent the energy of photon and light intensity, respectively. This time-periodic CPL polarizes in the 2D plane (i.e., the $x-y$ plane) and propagates along the $z$ (or $-z$) direction. As depicted in Fig. \ref{FIG1}(d), we show the photon-dressed band structures of VSi$_2$N$_4$ with a certain light intensity and frequency (i.e., $\hbar \omega=0.145$ eV and $eA/\hbar=0.028$ {\AA}$^{-1}$). In addition to the equilibrium bands (black solid lines), one can find that the Floquet-Bloch bands, that are created by absorbtion (blue dashed lines) or emission (red dashed lines) of a photon, are present. As expected, two Floquet-Bloch bands $E_c^{F}$ and $E_v^{F}$ simultaneously move close to the Fermi level. These Floquet-Bloch bands could be observed by utilizing time- and angle-resolved photoelectron spectroscopy (TR-ARPES) \cite{science.1239834,2016Selective,PhysRevX.10.041013}.  As shown in the insets of Fig. \ref{FIG1}(d), it is clearly found that two Floquet-Bloch bands (i.e., $E_c^{F}$ and $E_v^{F}$) invert to form the nontrivial band gap at the $K$ point while still preserve the trivial band gap at the $K'$ point. This indicates that VP-QAHE can be present in VSi$_2$N$_4$ subject to CPL. Besides, the results calculated from right-handed CPL only switch band topology between two valleys and exhibit the opposite propagating direction of chiral edge states (see Fig. S4 in the SM \cite{SM}). Thus, we mainly focus on the results of left-handed CPL in the following.

To elucidate topological phase transitions under irradiation of CPL, we further carefully examine light-induced evolution of Floquet-Bloch bands near the Fermi level. As shown in Figs.  \ref{FIG2}(a)-\ref{FIG2}(d) and Figs. \ref{FIG2}(e)-\ref{FIG2}(h), we respectively illustrate the evolution of Floquet-Bloch bands of VSi$_2$N$_4$ around the $K$ and $K'$ valleys at some typical photon energies with a fixed light intensity of $eA/\hbar$ = 0.032 {\AA}$^{-1}$.  With the increase of photon energy $\hbar \omega$, we can see that Floquet-Bloch bands $E_c^{F}$ and $E_v^{F}$ wholly tend to strengthen hybridization, but they exhibit different band topology near the $K$ and $K'$ valleys. As shown in Fig. \ref{FIG2}(i), we thoroughly check the band gaps near the $K$ and $K'$ points. The results indicate that band gap at the $K$ point closes and reopens twice [i.e., occurring at $\hbar \omega=0.142$ eV and $\hbar \omega=0.157$ eV, see Figs. \ref{FIG2}(a) and \ref{FIG2}(c)], while that at the $K'$ point closes and reopens once [i.e., occurring at $\hbar \omega=0.177$ eV, see Fig. \ref{FIG2}(g)]. More strikingly, the momentum of zero band gap away from high-symmetry point is more interesting [see Fig. \ref{FIG2}(c)], compared with the zero band gap located at the high-symmetry $K$ or $K'$ point [i.e., Figs. \ref{FIG2}(a) or \ref{FIG2}(g)]. Due to the threefold rotational symmetry around the $K$ points, a triangular distortion of the Fermi surface would be present, which is known as trigonal warping \cite{PhysRevLett.98.176806,PhysRevB.82.113405,PhysRevB.95.045424,PhysRevB.101.161103,PhysRevB.104.195427}. As shown in Fig. \ref{FIG2}(j), the energy difference map between Floquet-Bloch bands $E_c^{F}$ and $E_v^{F}$ in the vicinity of $K$ valley clearly indicates that there are three zero energy differences, which respectively locate at the $q_{1}$, $q_{2}$, and $q_{3}$ points. This confirms that the light-induced trigonal warping indeed arise in VSi$_2$N$_4$ subject to CPL. The presence of trigonal warping would like to strongly enrich topological phases of VP-QAHE and will be discussed in detail below.

\begin{figure}
    \centering
    \includegraphics[scale=0.98]{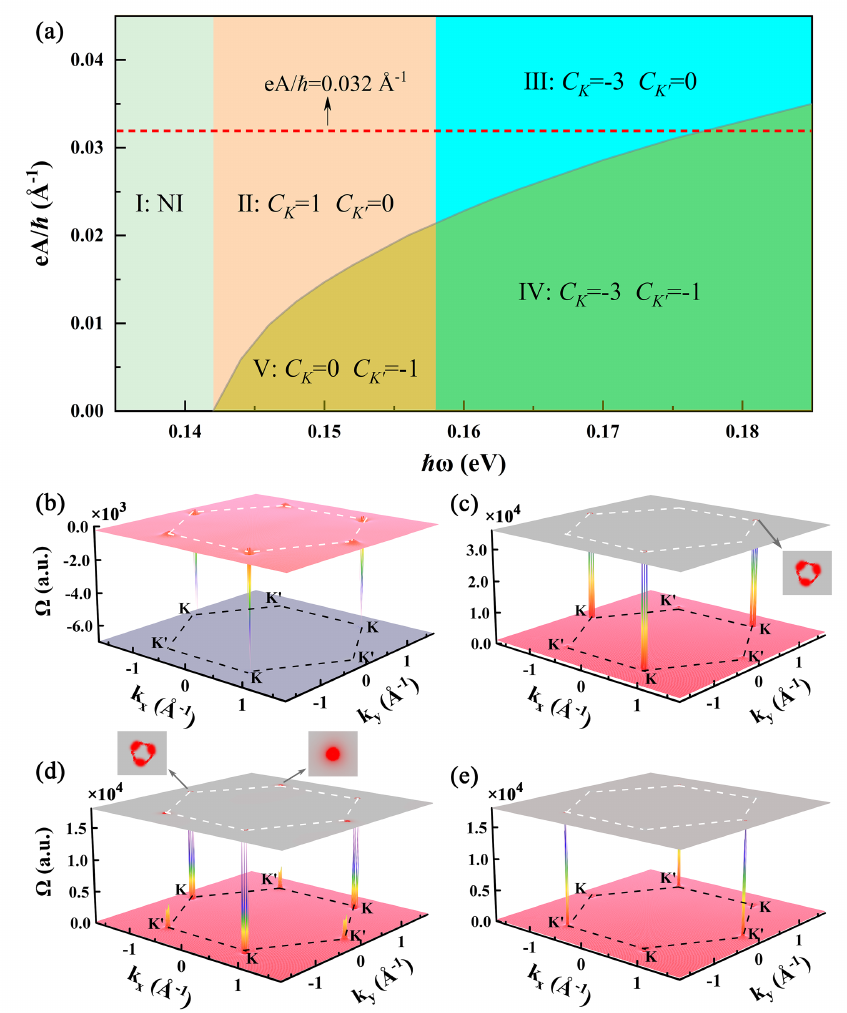}
    \caption{(a) The phase diagram as functions of $\hbar \omega$ and $eA/\hbar$. Five distinct topological phases, such as regime I: $\mathcal{C}_K = 0$ and $\mathcal{C}_{K'}=0$, regime II: $\mathcal{C}_K = +1$ and $\mathcal{C}_{K'}=0$, regime III: $\mathcal{C}_K = -3$ and $\mathcal{C}_{K'}=0$, regime IV: $\mathcal{C}_K = -3$ and $\mathcal{C}_{K'}=-1$, and regime V: $\mathcal{C}_K = 0$ and $\mathcal{C}_{K'}=-1$, are shown as different colors. The horizontal dashed line corresponds to the evolution with a fixed light intensity [see Fig. \ref{FIG2}(i)]. (b)-(e) The distribution of the Berry curvature $\Omega_z(\mathbf{k})$ under irradiation of left-handed CPL, which respectively correspond to the nontrivial regimes II, III, IV, and V in panel (a). Insets in panels (b) and (c) show the closeup of $\Omega_z(\mathbf{k})$ near the $K$ and $K'$ points, confirming the presence of trigonal warping near the $K$ valley.
    \label{FIG3}}
\end{figure}

Based on the above analysis, we can conclude that there are rich valley-related topological phases related to the VP-QAHE in VSi$_2$N$_4$ subject to CPL. To identify these distinct topological phases, it is required to calculate topological invariants by integrating the Berry curvature over the first BZ. Besides, the light intensity can also renormalize photon-dressed band structures and thereby affect the topological phase transition (see Fig. S2 in the SM \cite{SM}). To fully clarify coherent interactions between the frequency and intensity, we present a comprehensive study on Floquet Engineering of topological phases in VSi$_2$N$_4$. Since light-induced band inversion only occurs near the the $K$ and/or $K'$ valleys, we can well define the valley-resoled Chern numbers such as $\mathcal{C}_K$ and $\mathcal{C}_{K'}$, and then the global topological invariant (i.e., the first Chern number $\mathcal{C}$) is $C=\mathcal{C}_K + \mathcal{C}_{K'}$ \cite{PhysRevLett.112.106802,PhysRevLett.119.046403}. According to these topological invariants, we can completely establish the evolution of topological phases under irradiation of CPL.

\begin{figure}
    \centering
    \includegraphics[scale=1.03]{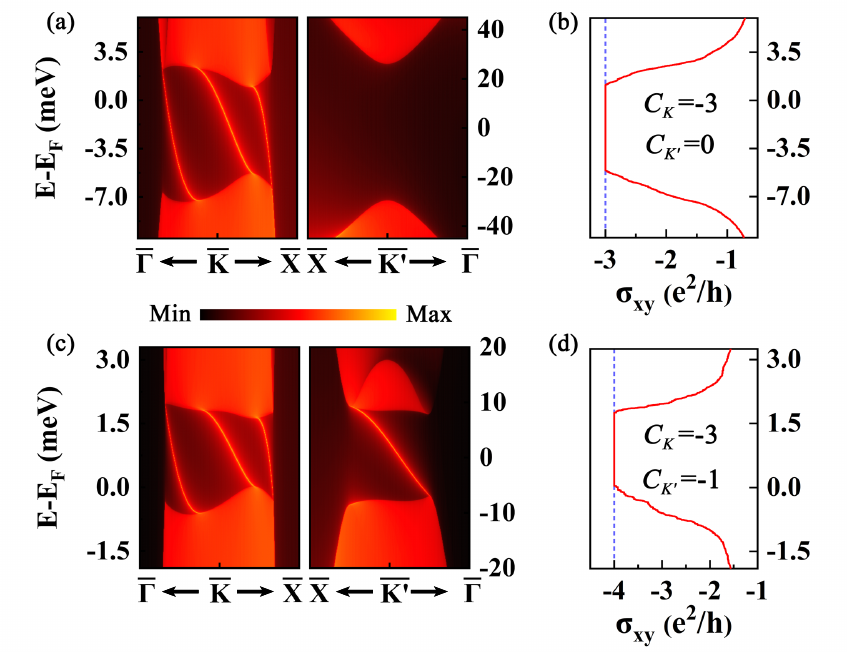}
    \caption{ (a) and (c) The calculated semi-infinite LDOS, and (b) and (d) anomalous Hall conductance under irradiation of left-handed CPL with photon energy $\hbar \omega = 0.185$ eV. In panels (a) and (b), the light intensity is set as 0.050 {\AA}$^{-1}$; in panels (c) and (d) the light intensity is set as 0.025 {\AA}$^{-1}$. The Hall conductances inside the nontrivial band gap are exactly quantized to $-3\textit e^{2}/h$ and $-4\textit e^{2}/h$ in panels (b) and (d), respectively.
    \label{FIG4}}
\end{figure}

As shown in Fig. \ref{FIG3}(a), the phase diagram characterized by $\mathcal{C}_{K (K')}$ as functions of $\hbar \omega$ and $eA/\hbar$ indicates that there are five distinct topological phases, such as regime I: $\mathcal{C}_K = 0$ and $\mathcal{C}_{K'}=0$, regime II: $\mathcal{C}_K = +1$ and $\mathcal{C}_{K'}=0$, regime III: $\mathcal{C}_K = -3$ and $\mathcal{C}_{K'}=0$, regime IV: $\mathcal{C}_K = -3$ and $\mathcal{C}_{K'}=-1$, and regime V: $\mathcal{C}_K = 0$ and $\mathcal{C}_{K'}=-1$. Except the topologically trivial regime I, other four regimes are all related to the topologically nontrivial VP-QAHE, and the corresponding distributions of Berry curvature are plotted in Figs. \ref{FIG3}(b)-\ref{FIG3}(e), respectively. As expected, the nonzero Berry curvature $\Omega_z(\mathbf{k})$ diverges near the $K$ and/or $K'$ points. As shown in the insets of Figs. \ref{FIG3}(c) and \ref{FIG3}(d), we can see that $\Omega_z(\mathbf{k})$ near the $K$ point are with respect to the threefold rotational symmetry, further confirming nontrivial band topology associated with trigonal warping. Significantly, the occurrence of trigonal warping is responsible for the phase transition from regime II to regime III [see the horizontal dashed line in Fig. \ref{FIG3}(a), which also corresponds to the process in Fig. \ref{FIG2}(i)], i.e.,  $\mathcal{C}_K = +1$ jumping to $\mathcal{C}_K = -3$; that is, trigonal warping plays a key role on the VP-QAHE with tunable Chern number.
%it is worth noting that $\mathcal{C}_K = +1$ jumps to $\mathcal{C}_K = -3$ form regime II to regime III [see the horizontal dashed line in Fig. \ref{FIG3}(a), which also corresponds to the process in Fig. \ref{FIG2}(i)], revealing that trigonal warping plays a key role on the VP-QAHE with tunable Chern number. %Moreover, topological phase transitions induced by tuning the light intensity with a fixed frequency is also visible; for instance, the transition between regimes II and V occurs along the vertical dashed line in Fig. \ref{FIG3}(a).
Consider that the change of handedness of CPL switches the chirality of edge channels (see Figs. S3-S5 in the SM \cite{SM}), the Chern number in VSi$_2$N$_4$ subject to CPL possesses a wide range as $\mathcal{C}= \pm1$, $\mathcal{C}= \pm3$, and $\mathcal{C}= \pm4$.

The Floquet VP-QAHE with specific first Chern number $\mathcal{C}$ and valley-resolved Chern number $\mathcal{C}_{K (K')}$ corresponds to the valley-dependent chiral edge channels and quantized Hall conductance $\sigma_{xy}$. To intuitively illustrate this, we calculate the local density of states (LDOS) based on the iterative Green's method \cite{Sancho_1985, WU2017} and the anomalous Hall conductance via the Kubo formula using the Floquet WFTB Hamiltonian \cite{SM}. Figures \ref{FIG4}(a) and \ref{FIG4}(c) show the light-induced LDOS of a semi-infinite ribbon of VSi$_2$N$_4$ with high Chern numbers such as $\mathcal{C}=-3$ ($\mathcal{C}_K = -3$ and $\mathcal{C}_{K'}=0$) and $\mathcal{C}=-4$ (i.e., $\mathcal{C}_K = -3$ and $\mathcal{C}_{K'}=-1$), respectively; the LDOS of low Chern numbers are shown in Fig. S4 of the SM \cite{SM}. The figures confirm that the chiral edge states are strongly valley-dependent, providing attractive Floquet Engineering of topological valleytronics based on light-induced VP-QAHE [as schematically plotted in Fig. \ref{FIG1}(b)]. In comparison with valley-resolved chiral edge channels, the intrinsic Hall conductances are exactly quantized to $-3\textit e^{2}/h$ and $-4\textit e^{2}/h$ inside the whole band gaps [see Figs. \ref{FIG4}(b) and \ref{FIG4}(d)], characterizing the global band topology of VP-QAHE.

In summary, we theoretically proposed a strategy beyond the existing paradigm to realize QAHE in nonequilibrium via hybridization of Floquet-Bloch bands. Combining with first-principles calculations and Floquet theorem, we found that our proposal of VP-QAHE can widely be realized in FM and NM candidates of 2D family materials $M$Si$_2$Z$_4$ ($M$ = Mo, W, V; $Z$ = N, P, As) under irradiation of CPL. Strikingly, light-induced trigonal warping and multiple band inversion are responsible that the Chern number of VP-QAHE is highly tunable and up to $\mathcal{C}=\pm 4$, which benefits to design high-performance QAHE devices. Considering that the light intensity and frequency, which induce topological phase transitions as shown in Fig. \ref{FIG3}(a), have been maturely employed to investigate various light-matter interactions in experiments \cite{science.1239834,mciver2020light,nanolett.6b04419,Light2022}, and even some candidates of 2D family materials $M$Si$_2$$Z_4$  were successfully synthesized \cite{science.abb7023}, our findings will stimulate the great experimental interest. Furthermore, the presence of VP-QAHE here is independent of magnetic order; thus, our proposal would provide a blueprint for systematic exploration of nonequilibrium QAHE in not only realistic materials but also artificial lattice systems.
%Our work offers a fascinating avenue to explore emergent nonequilibrium topological phases under light irradiation.

%in experimentally accessible regimes

This work was supported by the National Natural Science Foundation of China (NSFC, Grants No. 12222402, No. 11974062, and No. 12147102), the Shenzhen Institute for Quantum Science and Engineering (Grant No. SIQSE202101), and the Graduate Scientific Research and Innovation Foundation of Chongqing of China (Grant No. CYB22046).

%\bibliographystyle{apsrev4-2}
%\bibliography{ref}

%apsrev4-2.bst 2019-01-14 (MD) hand-edited version of apsrev4-1.bst
%Control: key (0)
%Control: author (72) initials jnrlst
%Control: editor formatted (1) identically to author
%Control: production of article title (-1) disabled
%Control: page (0) single
%Control: year (1) truncated
%Control: production of eprint (0) enabled
%

\end{document}